\documentclass[
  aps,
  prd,
  reprint,
  superscriptaddress,
  nofootinbib,
  nobibnotes,
  floatfix,
  amsmath,
  amssymb,
  showkeys
]{revtex4-2}

\usepackage[utf8]{inputenc}
\usepackage[T1]{fontenc}
\usepackage{graphicx}
\usepackage{amsfonts}
\usepackage{xcolor}
\usepackage{microtype}
\usepackage{hyperref}

\hypersetup{
    colorlinks=true,
    linkcolor=blue,
    citecolor=blue,
    filecolor=blue,
    urlcolor=blue,
    pdftitle={Extending the Pulsar radio population through Quantum Vacuum Cherenkov Emission},
    pdfauthor={Tales A. O. Gomes et al.}
}

\newcommand{\affilUFES}{N\'ucleo de Astrof\'isica e Cosmologia (Cosmo-Ufes) \& Departamento de F\'isica, Universidade Federal do Esp\'irito Santo, Av. Fernando Ferrari 514, Vit\'oria, 29075-910, ES, Brazil}
\newcommand{\affilELI}{Extreme Light Infrastructure ERIC, ELI Beamlines Facility, Za Radnic\'i 835, Doln\'i B\v{r}e\v{z}any, 25241, Czech Republic}
\newcommand{\affilINPE}{Divis\~ao de Astrof\'isica, Instituto Nacional de Pesquisas Espaciais, Avenida dos Astronautas 1758, S\~ao Jos\'e dos Campos, 12227-010, SP, Brazil}
\newcommand{\affilICRANet}{ICRANet, Piazza della Repubblica 10, Pescara, I-65122, Italy}
\newcommand{\affilICRANetFe}{ICRANet-Ferrara, Dipartimento di Fisica e Scienze della Terra, Universit\`a degli Studi di Ferrara, Via Saragat 1, Ferrara, I-44122, Italy}
\newcommand{\affilICRA}{ICRA, Dipartimento di Fisica, Sapienza Universit\`a di Roma, Piazzale Aldo Moro 5, Rome, I-00185, Italy}
\newcommand{\affilUniFe}{Dipartimento di Fisica e Scienze della Terra, Universit\`a degli Studi di Ferrara, Via Saragat 1, Ferrara, I-44122, Italy}
\newcommand{\affilINAF}{Istituto di Astrofisica e Planetologia Spaziali, INAF, Via Fosso del Cavaliere 100, Rome, I-00133, Italy}
\newcommand{\affilCTU}{Faculty of Nuclear Sciences and Physical Engineering, Czech Technical University in Prague, Brehova 7, Prague, 115 19, Czech Republic}

\begin{document}

\title{Extending the Pulsar radio population through Quantum Vacuum Cherenkov Emission}

\author{Tales A.~O. Gomes}
\email{tales.gomes@edu.ufes.br}
\affiliation{\affilUFES}

\author{Sergei~V. Bulanov}
\affiliation{\affilELI}

\author{Jaziel~G. Coelho}
\affiliation{\affilUFES}
\affiliation{\affilINPE}

\author{Martin Jirka}
\affiliation{\affilELI}
\affiliation{\affilCTU}

\author{Alexander~J. Macleod}
\affiliation{\affilELI}

\author{J.~A. Rueda}
\affiliation{\affilICRANet}
\affiliation{\affilICRANetFe}
\affiliation{\affilICRA}
\affiliation{\affilUniFe}
\affiliation{\affilINAF}

\date{\today}

\begin{abstract}
The presence of some radio pulsars within and below the so-called ``death valley'' in the $P$--$\dot{P}$ diagram challenges the standard pair-cascade models based on curvature radiation (CR) and resonant inverse Compton scattering (RICS) in neutron star polar caps. We show that Cherenkov radiation from the quantum vacuum, a strong-field quantum electrodynamics (QED) phenomenon enabled by vacuum polarization, constitutes a distinct pair-creation channel within the vacuum gap model. Within this model, the acceleration length scale for Cherenkov photon emission and the subsequent photon mean free path are sufficiently short to ensure rapid conversion into $e^+e^-$ pairs. This process acts as a complementary pair-creation channel, remaining viable even in regimes where classical mechanisms become heavily suppressed. This leads to an additional set of death lines in the $P$--$\dot{P}$ diagram that partially overlap with and extend beyond the CR and RICS death valleys, enlarging the viable parameter space for radio-emitting neutron stars. The result provides a physical basis for coherent radio emission from long-period neutron stars and offers an observable window for testing non-linear QED with next-generation radio facilities such as the Square Kilometre Array.
\end{abstract}


\maketitle

\section{Introduction}

The coherent radio emission from pulsars is a defining characteristic of these objects, commonly explained by electron-positron pair-creation cascades in the polar cap of a neutron star (NS). This pair-creation mechanism is commonly attributed to the interaction between high-energy photons and the pulsar's magnetic field \cite{Sturrock1971,Ruderman1975,philipov2020}. The condition on the pulsar parameter space for pair-creation defines the theoretical ``death line'', commonly translated into a boundary in the $P$-$\dot{P}$ diagram below which radio emission is expected to switch off. The region between the death line for pure dipolar and multipolar and highly-twisted fields forms a ``death valley'' where radio emission is predicted to suppress or cease entirely \cite{chen1993pulsar}. Otherwise, these lines can also define an active zone, a boundary which radio emission is expected to turn on \cite{cooper2024}.

Coherent radio emission produced close to the surface of pulsars is thought to stem from non-stationary pair-creation across acceleration gaps in the Timokhin-Arons mechanism \cite{timokhin2010,timokhin2013}. In this mechanism, leptons are accelerated across a vacuum gap ($E_{\parallel} \neq 0$) and produce high energetic photons, which decay to form a pair formation front (which screens the gap) at a distance $\sim h_{\text{gap}}$. The gap height depends on three length scales: the acceleration scale $\ell_{\rm acc}$ in which the primary charged particles acquire the Lorentz factor $\gamma_c$ sufficient to emit a photon of energy $\mathcal{E}_c$; the mean free path $\ell_{\rm em}$ over which the charged particle with $\gamma_c$ emits the photon via a given radiation process; and the mean free path $\ell_{\rm ph}$ of the photon before decaying into pairs via one-photon pair production ($\gamma+B\rightarrow e^-+e^+$). In this sense, the gap height is given by $h_{\rm gap} = \ell_{\rm acc} + \ell_{\rm ph}$, as long as the mean free path of emission is much smaller than the gap height, $\ell_{\rm em}\ll h_{\rm gap} $. The ``pulsar death lines'' \cite{Ruderman1975} mark the threshold where the kinematics conditions for the pair cascade to occur is satisfied, and enough pairs are produced to screen $\vec{E}_\parallel$, leading to the cessation of radio emission.

However, this classical paradigm has been challenged by the discovery of rotationally-powered radio pulsars located deep within or beyond the death valley, such as PSR J0250+5854 ($P = 23.5$~s) \cite{Tan_2018} and the ultra-long-period pulsar PSR J0901$-$4046 ($P = 75.9$~s) \cite{caleb2022}. For these sources, the available spin-down energy remains sufficient to power the observed radio luminosity, yet standard curvature radiation polar-cap mechanisms fail to operate at such slow rotation rates. While pair cascades might theoretically be sustained through Resonant Inverse Compton Scattering (RICS), this mechanism stringently requires the presence of complex multipolar surface magnetic fields (with local curvature radii $\rho_c \sim 10^5$~cm) and high surface temperatures to enclose these sources within an active zone \cite{cooper2024}. Concurrently, the recent discoveries of long-period radio transients (LPTs, e.g., \cite{wang2025,2023Natur.619..487H}) reveal an even more distinct population of slowly rotating compact objects, though these are generally understood to be magnetically-powered systems or magnetic white dwarf binaries rather than rotationally-powered neutron stars \cite{thompson1996,2024NatAs...8.1159C,katz2022,loeb2022,rea2025magnetars,rea2026}. Focusing on rotationally-powered pulsars in tension with theoretical death lines, a revised pair-creation framework grounded in first-principles QED is essential to provide insight into radio-emitting pulsars, without necessarily invoking highly localized magnetic anomalies. In this work, we focus on the electrostatic structure of the vacuum gap (VG) model, assuming a strong binding of charges at the stellar surface that allows a substantial parallel electric field to develop. 

Strong-field QED provides a compelling framework for such a review. The Schwinger critical magnetic field $B_S = m_e^2 c^3/e \hbar \approx 4.41 \times 10^{13}$~G (with an equivalent electric field $E_S = B_S$) marks the scale where QED becomes nonlinear. Fields approaching or exceeding $B_S$ are found at the surfaces of magnetars, which have dipolar surface fields of $B \sim 10^{14}$--$10^{15}$~G~\cite{mcgill2014, rea2025magnetars}, and near the surfaces of a small subset of high-field radio pulsars. For the broader pulsar population, surface fields typically range from $10^{11}$ to $10^{13}$~G, reaching up to a few times $B_S$ only in the most strongly magnetized objects. In such highly magnetized environment, the QED vacuum may act as a polarized medium, reducing the phase velocity of light \cite{EulerHeisenberg1936, toll1952, ADLER1971}, making possible the emission of Cherenkov radiation from the quantum vacuum \cite{Macleod2019}. Although such fields remain beyond terrestrial laboratory capabilities, next-generation facilities such as ELI \cite{ELIBeamlinesFacility}, APOLLON \cite{APOLLON}, NSF-OPAL \cite{NSFOPAL}, and SEL \cite{SEL} are planning to work in the regimes where strong-field QED effects come into play \cite{BET2003,MTB2006}. Strong-field QED has been demonstrated conclusively in experiments such as the SLAC E144, which observed nonlinear Compton scattering and multiphoton Breit-Wheeler pair-creation (one-photon pair production) by colliding $46$~GeV electrons with intense laser pulses \cite{Bamber1996,bula1996,Burke1998}. The ultra-peripheral heavy-ion collisions at the LHC have revealed light-by-light scattering in the strong electromagnetic fields of relativistic nuclei \cite{Aaboud2017,atlas2019}. The PVLAS experiment has set the current best limits on vacuum magnetic birefringence and dichroism \cite{EJLLI20201}. These results have experimentally verified strong-field QED effects, motivating their application to pulsar magnetospheres.

In this work, we show that Cherenkov radiation from the quantum vacuum constitutes a possible pair-creation channel in neutron star polar caps. To establish this, we first detail the theoretical methods, calculating the required electrodynamic gap heights for cascade formation across different emission mechanisms. Subsequently, we present the results by mapping the theoretical boundaries for coherent radio emission in the $P$-$\dot{P}$ diagram, including the death lines for curvature radiation and the active zone for RICS. Within this framework, we show that the acceleration length scale for Cherenkov photon emission and the subsequent photon mean free path are sufficiently short to ensure rapid conversion into $e^+e^-$ pairs. Consequently, this quantum Cherenkov (QC) mechanism acts as a complementary pair-creation channel that remains viable even in regimes where classical mechanisms become heavily suppressed or need a more complex field configuration. Our analysis leads to an additional set of death lines in the $P$-$\dot{P}$ diagram that partially overlap with and extend beyond the classical curvature radiation and RICS death valleys, thereby enlarging the viable parameter space for radio-emitting neutron stars.

\section{Methods}\label{sec6}

Pulsars are understood to be rotating NSs in which, due to their intense magnetization and superconducting interiors, external rotating dipole magnetic fields and quadrupole electric fields are generated. These objects form a magnetosphere with particle number density $n_{\rm GJ} \simeq \Omega B/(2\pi e c)\approx 3.1\times10^{13}P^{-2}B_*$ cm$^{-3}$ \cite{GoldreichJulian1969}, where ${B}$ is the dipolar magnetic field (with $B_* = B/B_S$) and ${\Omega} = 2\pi/P$ is the rotation angular frequency. This magnetosphere corotates rigidly with the NS up to the light cylinder radius, $R_{\rm lc} = c/\Omega \approx 4.8 \times 10^9 P$ cm, defined as the distance from the rotation axis where field lines would theoretically exceed the speed of light. At the poles, the last closed field lines originate from the polar cap region, which is characterized by an opening angle $\theta_{\rm pc} \approx \sqrt{R/R_{\rm lc}} \sim 1.5 \times 10^{-2} P^{-1/2}$ and a radius $R_{\rm pc} = R \theta_{\rm pc} \sim 1.65 \times 10^{4} P^{-1/2}$ cm, assuming a stellar radius of $R \approx 1.1 \times 10^6$ cm. The surface magnetic field at the pole for a dipolar configuration is $B= 6.4\times10^{19}\sqrt{P\dot{P}}$ G \cite{shapiro83}, resulting from equating the magnetic dipole luminosity to the NS rotational energy loss, adopting the above-mentioned radius and $10^{45}$ g cm$^2$ as a fiducial value for the moment of inertia.

The electrostatic structure of these acceleration regions and the screening of $E_\parallel$ depend fundamentally on the physical conditions at the stellar surface. Here, we focus on the VG model, where a vacuum region develops between the surface of the NS and the magnetosphere, allowing $\vec{E}_\parallel$ to grow \cite{Ruderman1975}. Within this model, the potential is the solution of the one-dimensional Poisson's equation, $\Delta V = (B\Omega/c)\ell_{\rm acc}^2$. The maximum available potential is the difference between the pole and the edge of the polar cap at the stellar surface, $\Delta \Phi_{\rm max}\simeq B R^3 \Omega^2/2c^2$. Comparing the acceleration potential drop with this maximum potential yields the maximum acceleration length scale $\ell_{\rm acc}^{\rm max}\approx R_{\rm pc}/\sqrt{2}$.

One-photon pair-creation is a fundamental process in pulsar magnetospheres and electron-laser collisions within a strong background electromagnetic field, where the key parameter is $\chi_\gamma = (\mathcal{E}_c/m_e c^2)(B_{\perp,\gamma}/B_S)$, with $\mathcal{E}_c$ the characteristic energy of the propagating photon and $B_{\perp,\gamma}=B \sin\theta_{\perp,\gamma}$ the transverse component of the magnetic field with respect to the photon propagation. The photon decay probability increases exponentially for $\chi_\gamma \gtrsim 1/15$ \cite{Erber1966,chen1993pulsar}. Consequently, photons emitted parallel to the magnetic field only decay after acquiring a sufficient perpendicular component $B_{\perp,\gamma}\sim (\ell_{\rm ph}/\rho_c) B$, defining the photon mean free path $\ell_{\rm ph} = \frac{m_e c^2}{15 \mathcal{E}_c}\frac{B_S}{B} \rho_c$. 

Along with $\ell_{\rm ph}$, the cascade dynamics in pulsars are governed by a set of critical length scales \cite{Hibschman_2001,cooper2024}: the free acceleration length $\ell_{\rm acc}$, the primary particle energy-loss scale $\ell_{\rm loss}$ for a given channel, the emission length scale $\ell_{\rm em}$ for a photon of energy $\mathcal{E}_c$, and the photon splitting scale ($\gamma\rightarrow\gamma\gamma$) $\ell_{\rm split}$. A successful cascade requires the hierarchy $\ell_{\rm loss} > h_{\rm gap} = \ell_{\rm acc} + \ell_{\rm ph} \gg \ell_{\rm em}$ and $\ell_{\rm ph} \ll \ell_{\rm split}$. This ensures that particles remain in the free acceleration regime, emit photons, and these photons decay into pairs before splitting.

The physical limit of this cascade is reached when the required gap height exceeds the available acceleration zone. Therefore, equating the required gap height to the maximum available acceleration length ($h_{\rm gap} \sim \ell_{\rm acc}^{\rm max}$) defines the theoretical ``pulsar death lines'', beyond which primary particles can no longer sustain the cascade, and radio emission ceases. Specifically, RICS constraints define a closed ``active zone'' for coherent emission, whereas QC and CR establish strict death lines bounding a ``death valley''.

In this section, we detail the analytical derivations of the polar gap height ($h_{\rm gap}$) and the corresponding acceleration potential ($\Delta V$) required to sustain electron-positron pair cascades for the three primary radiation mechanisms discussed in this work: Curvature Radiation (CR), Resonant Inverse Compton Scattering (RICS), and Quantum Cherenkov (QC) emission.

\subsection{Curvature radiation gap}
The photons emitted via CR have a characteristic energy $\mathcal{E}_c^{\rm CR} = (3/2)(\hbar c/\rho_c)\gamma_c^3$, where $\gamma_c\approx e\Delta V/(m_e c^2)$ is the Lorentz factor of the electron and $\rho_c$ is the radius of curvature of the magnetic field. For a pure dipole, we approximate the average curvature as $\rho_c \simeq \sqrt{R R_{\rm lc}} \sim 7\times10^7 \,P^{1/2}$~cm, whereas for a multipolar configuration, we adopt $\rho_c \sim R$. The total radiated power is $\mathcal{P}_{\rm CR} = (2/3)(\alpha \hbar c^2 / \rho_c^2) \gamma_c^4$, implying the emission length scale $\ell_{\rm em} \simeq c \mathcal{E}_c^{\rm CR} / \mathcal{P}_{\rm CR} = 9\rho_c / (4\alpha \gamma_c)$, where $\alpha = e^2/(\hbar c)$ is the fine-structure constant. Since CR photons are emitted almost tangentially to the field lines, they must travel a distance $\ell_{\rm ph}$ to acquire a sufficient perpendicular component $B_{\perp,\gamma} \approx (\ell_{\rm ph}/\rho_c)B$ to trigger pair-creation. Thus, the photon mean free path can be written as $\ell_{\rm ph} \approx 2 \rho_c^2 / (45 \lambda_c \gamma_c^3 B_*)$, where $\lambda_c = \hbar /(m_e c)$ is the electron Compton wavelength. Now, we find the CR gap height by finding the minimum of $h_{\rm gap}$ with respect to $\ell_{\rm acc}$, and approximating $h_{\rm gap}\approx \ell_{\rm acc}$, thus
\begin{equation}
    h_{\rm gap, CR} = 5.7\times10^2\,\,P^{3/7}\,{B_*}^{-4/7}\rho_6^{2/7}\,\,\text{cm},
\end{equation}
and the corresponding potential drop
\begin{equation}
    \Delta V_{\rm CR} = 9.1 \times 10^{11}\,\,P^{-1/7}\,{B_*}^{-1/7}\rho_6^{4/7}\,\,\text{V},
\end{equation}
where $\rho_6$ is the curvature radius $\rho_c$ in units of $10^6$~cm.

\subsection{Resonant Inverse Compton Scattering gap}

When soft surface thermal X-ray photons are present, primary electrons accelerated by a potential difference $\Delta V$ can experience catastrophic energy losses via Resonant Inverse Compton Scattering (RICS). As detailed in \cite{cooper2024}, for this channel to govern the pair cascade, the free acceleration length must match the RICS energy-loss scale ($\ell_{\rm acc} \approx \ell_{\rm loss, RICS}$) before other radiative processes intervene. Because the target photons constitute a thermal distribution with a variety of incoming angles, this resonant cooling ``wall'' is extensive and robustly accessed without fine-tuning. 

The primary particles are accelerated until they encounter this resonant wall at a critical Lorentz factor
\begin{equation}
    \gamma_{\rm int} \approx \frac{2 B_*}{3\Theta|W_{-1}(\mathcal{Z}_{\rm rot})|},
\end{equation}
where $B_* = B/B_S$ and $\Theta = k_B T / (m_e c^2)$ is the dimensionless surface temperature \cite{baring2011}. The length scale for a primary electron to reach this Lorentz factor is
\begin{equation}
    \ell_{\rm int} \approx\left[\frac{\lambda_c c P}{3\pi\Theta|W_{-1}(\mathcal{Z}_{\rm rot})|}\right]^{1/2}.
\end{equation}
The scattering kinematics are governed by the dimensionless argument $\mathcal{Z}_{\rm rot}$. 
After the primary electrons reach the critical Lorentz factor $\gamma_e\sim\gamma_{\rm int}$, RICS photons are emitted with a characteristic energy $\mathcal{E}_{\rm RICS} \approx 2\gamma_{\rm int}\,m_ec^2B_*/(1+2B_*)$, which gives the photon mean free path
\begin{equation}
    \ell_{\rm ph} \approx 3\rho_c\frac{(1+B_*)}{2B_*^2} | W_{-1}(\mathcal{Z}_{\rm rot})|.
\end{equation}
The total gap height here is $h_{\rm gap, RICS} \approx \ell_{\rm int} + \ell_{\rm ph}$, however, in the strong-field limit the RICS gap is predominantly set by the photon mean free path, which yields $h_{\rm gap, RICS} \approx 30 \rho_c \Theta / B_*$. Normalizing the radius of curvature as $\rho_6 = \rho_c/10^6\text{ cm}$, the gap height is explicitly expressed as
\begin{equation}
    h_{\rm gap, RICS} \approx 5.06 \times 10^3 B_*^{-1} \rho_6 T_6 \text{ cm},
\end{equation}
and the corresponding potential drop
\begin{equation}
    \Delta V_{\rm RICS} \approx 7.1 \times 10^{13} B_*^{-1} P^{-1} \rho_6^2 T_6^2 \text{ V},
\end{equation}
where $T_6 = T/(10^6\,\rm K)$ is the normalized surface temperature of the star.

\subsection{Quantum Cherenkov gap}
In QED, the possible interaction of photons and the background magnetic field leads to the fact that the field plays the role of a homogeneous anisotropic medium with dispersion and absorption, governed by the quantum parameter $\chi_\gamma$. The refractive index of such a medium can be defined in terms of the photon mass operator $n_{\parallel,\perp}^2-1 =-\mu_{\parallel,\perp}^2 c^4 / \hbar^2\omega^2$. In particular, in the weakly nonlinear regime $\chi_\gamma \ll 1$, we have
\begin{equation}
    \mu_{\pm}^2 = -\alpha m_e^2 \left[ \frac{11 \mp 3}{90\pi}\chi_\gamma^2 + i\sqrt{\frac{3}{2}}\frac{3 \mp 1}{16}\chi_\gamma e^{-8/3\chi_\gamma} \right].
\end{equation}
Here, the real part of $\mu_{\parallel,\perp}^2$ modifies the dispersion relation, whereas its imaginary part determines the pair-creation rate. For $\chi_{\gamma}\gtrsim1$ the imaginary component grows significantly faster than the real part, rendering one-photon pair production the dominant process.

Focusing on the real, dispersive contribution in the limit $\chi_\gamma \ll 1$, where $\Delta n_{\parallel,\perp} \equiv n_{\parallel,\perp}-1\ll 1$, substituting the real part of $\mu_\pm^2$ into the dispersion relation yields the refractive index deviation
\begin{equation}
\label{indexref}
    \Delta n_{\parallel,\perp}
    \approx \frac{11\mp 3}{180\pi}\,\alpha\left(\frac{B_{\perp,e}}{B_S}\right)^2
\end{equation}
with $B_{\perp,e}=B\sin\theta_{\perp,e}$ the perpendicular component of the magnetic field with respect to the $e^\pm$ trajectory at the emission of the QC photon.

One consequence of a refractive index greater than one is the emission of QC photons. The quantum vacuum supports two distinct polarization modes where the refractive indices differ by a factor of $\Delta n_\perp\sim1.75\Delta n _\parallel$. However, this difference translates to a threshold variation of $\gamma_{\rm Che,\parallel}\sim0.76~\gamma_{\rm Che,\perp}$, leading to only a marginal difference in the logarithmic scale of the resulting death lines. According to Cherenkov radiation theory \cite{Frank1937}, and its extension to polarized quantum vacuum \cite{Macleod2019,Bulanov2019}, we can approximate the angle of emission of QC photons, the Cherenkov angle, and the power spectrum as 
\begin{equation*} 
    \sin^2\theta_{\text{Che}}^{\parallel,\perp}\approx{2\Delta n_{\parallel,\perp}}\quad\text{and}\quad
\end{equation*}
and
\begin{equation}
\label{ChemodelQV}
    \frac{d\mathcal{P}_{\text{Che}}}{d\mathcal{E}}\approx\frac{\alpha}{4\pi\hbar}\sin^2\theta_{\text{Che}}^{\parallel,\perp}\mathcal{E}.
\end{equation}
The refractive index has a small peak at $\chi_\gamma^{\rm Che}\approx0.75$ and drops exponentially above it \cite{RITUS1972,toll1952}; therefore, we assume that QC photons are emitted with characteristic quantum parameter $\chi_\gamma^{\rm Che} \approx 0.75$, which gives the characteristic energy $\mathcal{E}_c^{\rm Che} = \chi_\gamma^{\rm Che}\,m_e c^2 (B_S/B_{\perp,e})$. This ensures they are produced well above the pair-creation threshold, typically associated with $\chi_\gamma\approx1/15$. 

A natural condition that arises from the theory is that the particle must have a velocity greater than the phase velocity of light for the emission of Cherenkov photons, leading to a threshold on the Lorentz factor, $\gamma>\gamma_{\rm Che} \approx 1/\sqrt{2\Delta n_{\parallel,\perp}}\sim60\,(B_S/B_{\perp,e})$. From $\gamma_c = e\Delta V/2m_e c^2$, we have
\begin{equation}
    \label{eq:l_acc}
    \ell_{\rm acc} \approx \frac{4.7}{\sqrt{\sin\theta_{\perp,e}}}\,\,P^{1/2}B_*^{-1}\,\,\text{cm}.
\end{equation}
Since QC photons are characterized by a constant quantum parameter $\chi_\gamma^{\rm Che} \approx 0.75>1/15$ \cite{toll1952}, their mean free path will be very small compared to the acceleration length. Evaluating the exact QED attenuation coefficient \cite{Erber1966} at this peak yields the photon mean free path:
\begin{equation}
\begin{split}
    l_{\rm ph}(\chi_\gamma \approx 0.75) &\approx 22 \left( \frac{\lambda_c}{\alpha_f} \right) \left( \frac{B_S}{B_{\perp,e}} \right) \\[1ex]
    &\approx 1.1 \times 10^{-7} B_{*,\perp}^{-1} \text{ cm}.
\end{split}
\end{equation}
This microscopic conversion scale is many orders of magnitude smaller than the macroscopic acceleration length evaluated in Eq.~\eqref{eq:l_acc}. Therefore, the condition $l_{\rm ph} \ll l_{\rm acc}$ is satisfied, allowing us to safely approximate the total gap height as $h_{\rm QC} \approx l_{\rm acc}$.
Now, integrating \eqref{ChemodelQV} and using $\ell_{\rm em}\simeq c\mathcal{E}_c^{\rm Che}/\mathcal P_{\rm Che}$, we obtain the emission length
\begin{equation}
    \ell_{\rm em}\approx6.48\times10^{-4}\,\sin\theta_{\perp,e}^{-1}\,\,B_*^{-1}\,\,\text{cm}.
\end{equation}
Imposing $\ell_{\rm em} \ll \ell_{\rm acc}$ yields $P \gg 1.91\times10^{-8}\sin\theta_{\perp,e}^{-1}$. Conversely, requiring the acceleration length to remain strictly smaller than the polar cap radius ($\ell_{\rm acc} \ll R_{\rm pc}$) implies $P \ll 3.5\times10^3\sqrt{\sin\theta_{\perp,e}}B_*$. Combining these limits, we find that the magnetic field must satisfy $B_* > 2.97\times10^{-9}P^{3/4}$ for the mechanism to operate. For the vast majority of isolated pulsars, this condition is comfortably satisfied. Finally, we can write the gap height and the corresponding potential drop in the QC case as
\begin{equation}
    h_{\rm gap, QC} = \frac{4.7}{\sqrt{\sin\theta_{\perp,e}}}\,\,P^{1/2}B_*^{-1}\,\,\text{cm},
\end{equation}
\begin{equation}
    \Delta V_{\rm QC} = \frac{6.1\times10^7}{\sin\theta_{\perp,e}}\,B_*^{-1}\,\text{V.}
\end{equation}

\begin{figure*}[t]
    \centering
    \includegraphics[width=0.65\textwidth]{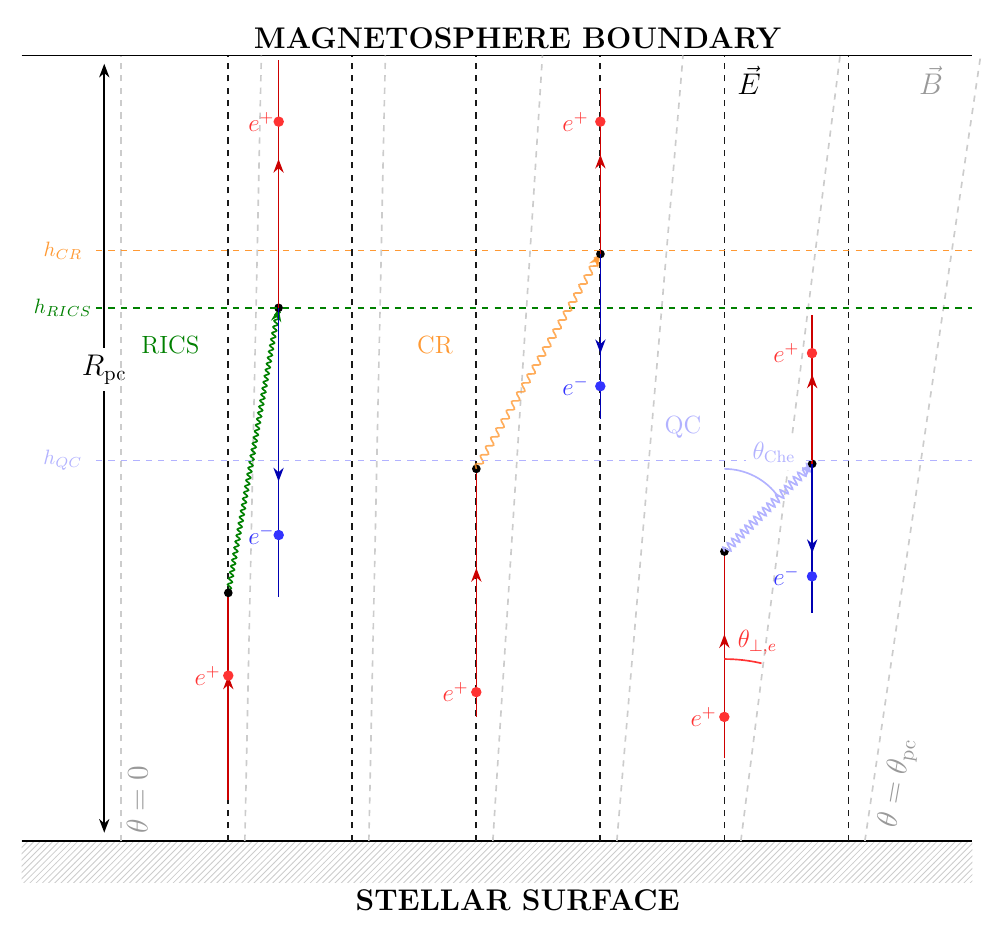}
    \caption{\textbf{Schematic representation of electron-positron ($e^\pm$) pair cascade formation in the pulsar polar cap gap.} The diagram illustrates the gap extending from the stellar surface to the magnetosphere boundary (maximum height $R_{\rm pc}$). Primary positrons ($e^+$, red circles) are accelerated by the vertical electric field $\vec{E}$ and follow the magnetic field lines $\vec{B}$ (light grey dashed lines), which diverge from the magnetic axis ($\theta = 0$) toward the polar cap edge ($\theta = \theta_{\rm pc}$). Three photon emission mechanisms and their subsequent decay into secondary pairs are illustrated: (i) Resonant Inverse Compton Scattering (RICS, green wavy lines), (ii) Curvature Radiation (CR, orange wavy lines), and (iii) Quantum vacuum Cherenkov emission (QC, blue wavy lines). For the QC process, the positron pitch angle $\theta_{\perp,e}$ and the characteristic Cherenkov emission angle $\theta_{\rm Che}$ are highlighted. The emitted gamma-ray photons (wavy lines) decay into secondary $e^+e^-$ pairs via one-photon pair production ($\gamma \to e^+e^-$), sustaining the magnetospheric plasma.}
    \label{fig:figura}
\end{figure*}

\section{Results}

To map the viable emission boundaries across the pulsar population in the $P$-$\dot{P}$ diagram, we evaluate the physical limit of the pair cascade for different radiative processes. In the vacuum gap model, this threshold is reached when the required gap height ($h_{\rm gap}$) equates to the maximum available acceleration length ($\ell_{\rm acc}^{\rm max}$). Applying this condition each process yields the following boundaries in the $P$--$\dot{P}$ diagram.

For standard Curvature Radiation (CR), the cascade conditions establish death lines bounding a ``death valley'', given by
\begin{align}
    \label{cond1a} \log{\dot{P}} &= \frac{11}{4}\log P - 14.62 , \\
    \label{cond1b} \log{\dot{P}} &= \frac{9}{4}\log P - 16.58 ,
\end{align}
for dipolar and multipolar field configurations, respectively. These lines define the ``death valley'' region.

For RICS, coherent radio emission in the $P-\dot{P}$ space is not bounded by a simple death line, but rather delineated by a closed ``active zone'' governed by three simultaneous physical constraints \cite{cooper2024} in the strong-field limit ($\mathcal{B} \gg 1$). First, thermodynamic gap viability imposes an absolute minimum rotation period that depends strictly on the stellar surface temperature, $P \gtrsim 120 \left(T/{10^{6.5}\text{ K}}\right)^{-5} \text{ s}$. Second, the cascade must achieve a critical pseudo-multiplicity ($\mathcal{M}_{\rm rot} \ge 1$), with $\mathcal{M}_{\rm rot}\propto \dot{P}^{-1/2}\,P^{-1}$ which yields $\dot{P}\propto P^{-2}$, guaranteeing sufficient pair plasma proliferation to supply the required screening current. Finally, the system must satisfy a energy budget: the power demanded by the accelerator to produce the primary pair population ($L_{e^+e^-}\approx e\Delta V_{\rm RICS} \dot{N}$) at the polar cap cannot exceed the available spin-down luminosity of the pulsar ($L_{\rm sd}\approx 4\pi^2I\dot{P}P^{-3}$). By equating these luminosities ($L_{e^+e^-} \le L_{\rm sd}$), the critical energy boundary for rotationally powered pulsars scales as $\dot{P} \propto P^{3/5}$, in the strong-field limit. The full analytical derivations of the gap height and potential drop are detailed in the Methods section.

The death lines for QC model are obtained similarly as CR model, the main difference is in the threshold in the Lorentz factor of the electron for the emission and the characteristic energy which ensures a rapid conversion into pairs. We assume that the electric field is predominantly radial, whereas the magnetic field can have either a dipolar or a highly curved configuration. Consequently, the charged particles accelerated by the electric field intersect the magnetic field lines at an angle approximating the maximum polar cap angle, $\theta_{\rm pc}$. Assuming $B_{\perp,e}=B\sin\theta_{\rm pc}$ for a dipolar configuration and $B_{\perp,e}=B$ for a multipolar configuration, the death lines for this model are given by
\begin{equation}
\label{cond3a}
    \log{\dot{P}} = \frac{3}{2}\log{P}-17.29,
\end{equation}
\begin{equation}
\label{cond3b}
    \log{\dot{P}} = \log{P} - 19.11\,.
\end{equation}
Since our analysis is performed using $\Delta n_{\parallel,\perp}\lesssim1$, our results are limited by $B\lesssim\frac{180\pi}{(11\pm3)\alpha} B_S\sim10^{17}$ G. Above this limit, the vacuum index of refraction deviates drastically from unity, breaking the weak dispersion assumptions underlying our analytical framework. 

\begin{figure*}[t]
\centering
\includegraphics[width=0.85\textwidth]{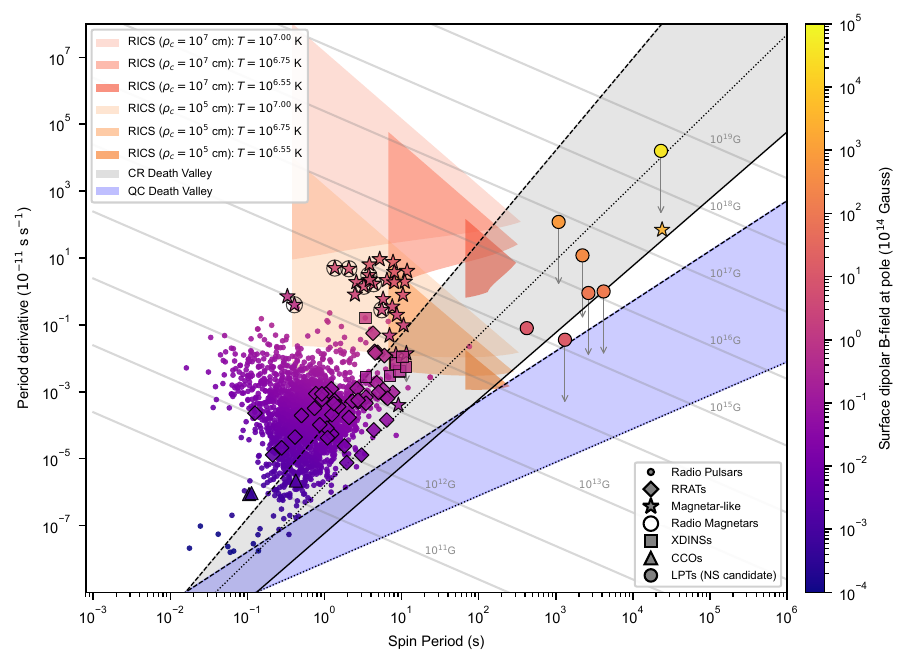}
\caption{\textbf{Pulsar period and period derivative diagram.} The diagram shows observed isolated neutron stars, magnetars, and long-period radio transients. Diamonds represent rotating radio transients, squares indicate X-ray dim isolated neutron stars, and upward-pointing triangles represent central compact objects. Small circular dots show normal radio pulsars from the ATNF catalog \cite{atnf}, and star symbols denote pulsars with magnetar-like emission \cite{rea2025magnetars,mcgill2014}, including the long-period magnetar 1E 161348-5055 \cite{deluca2006,rea2016}. Long-period radio transients (neutron star candidates) are represented as large circles \cite{rea2026}. The color of all data markers corresponds to the surface dipolar magnetic field at the pole, as indicated by the right-hand color bar. Vertical arrows represent period derivative upper limits. Oblique lines, forming the shaded areas, represent different pulsar death lines and active zones in the vacuum gap model \cite{Ruderman1975}. Gray shaded areas are for curvature radiation \cite{chen1993pulsar}, orange and red shaded areas are the active zones for resonant inverse Compton scattering \cite{cooper2024}, and the blue shaded area represents the quantum Cherenkov emission death valley [Eqs.~\eqref{cond3a} and \eqref{cond3b}]. This figure was generated making use of open-source code available at: \url{https://github.com/nhurleywalker/GPMTransient}.}
\label{fig:dlnew}
\end{figure*}

The resulting theoretical boundaries for CR, RICS, and QC are visualized in the $P-\dot{P}$ diagram (Fig.~\ref{fig:dlnew}) alongside the known population of isolated neutron stars. To contextualize these limits, the observational data includes normal radio pulsars, Rotating Radio Transients (RRATs), Central Compact Objects (CCOs), X-ray Dim Isolated Neutron Stars (XDINSs), magnetars, and the recently discovered long-period radio transients (LPTs). This comparison demonstrates how the QC mechanism enlarges the viable parameter space for coherent radio emission beyond the classical CR and RICS limits.

\section{Discussion}\label{sec4}

Radiation produced by accelerated charges (such as synchrotron, curvature, and inverse Compton) and Cherenkov radiation, which are usually treated as distinct emission processes, can be regarded as limiting regimes of a single underlying QED mechanism \cite{Schwinger1976,Bulanov2019}. Pulsars provide a natural environment for studying the interplay between these regimes, since both ultra-relativistic particle acceleration and strong-field vacuum polarization operate simultaneously. In the presence of intense magnetic fields, the quantum vacuum becomes polarized, modifying the photon dispersion, so enabling the QC onset in addition to classical radiation by acceleration \cite{Sauter1931,Frank1937,toll1952,schwinger1949,schwinger1951,narozhny1979,ritus1985quantum}.

The three pair-creation channels considered here operate on distinct characteristic length scales within the vacuum gap. CR photons are emitted on a short scale but must propagate a significant distance before the perpendicular component of the magnetic field is sufficient to trigger decay into pairs. RICS photons, produced via scattering of thermal polar-cap photons by relativistic electrons, are emitted on a longer scale but are more energetic and are absorbed faster. QC photons are produced through the interaction of $e^\pm$ with the polarized quantum vacuum and are only emitted once the particle exceeds the Cherenkov threshold $\gamma > \gamma_{\rm Che}$. The acceleration length scale for QC emission and the photon mean free path before pair-creation satisfy $\ell_{\rm ph} \ll h_{\rm QC}$, ensuring rapid conversion into $e^+e^-$ pairs within the gap. Rather than necessarily dominating over CR or RICS across the entire parameter space, QC acts as a robust complementary pair-creation channel. See Fig.~\ref{fig:figura} for a schematic representation.

These properties place the QC death lines in a distinct region of the $P$-$\dot{P}$ diagram (Fig.~\ref{fig:dlnew}), partially overlapping with and extending beyond the CR death valley and RICS active zone. Pair-creation via QC can therefore remain operative in regimes where the other two mechanisms become inefficient, sustaining the pair cascade and, consequently, allowing coherent radio emission in an enlarged region of pulsar parameter space. This constitutes the central result of the present work. The QC mechanism does not replace classical radiation (CR and RICS); rather, it complements it in the strong-field regime where the polarized vacuum becomes the primary radiative medium.

A natural question concerns the applicability of this framework to objects with fields as low as $B \sim 10^{11}$~G, where $\Delta n \sim 10^{-15}$ and the QC power is suppressed by a factor $B_*^2 \sim 10^{-11}$. Although this suppression is significant, the death line is not primarily determined by the radiated power, but by the self-consistency of the pair-creation chain: (i) acceleration of primary particles to $\gamma >\gamma_{\rm Che} \approx 60\,\theta_{\rm pc}^{-1}B_*^{-1}$ within the gap; (ii) emission of QC photons with characteristic quantum parameter $\chi_\gamma^{\rm Che} = 0.75$, well above the pair-creation threshold $\chi_\gamma \approx 1/15$; and (iii) photon absorption on a length scale $\ell_{\rm ph} \ll h_{\rm gap}$. Conditions (ii) and (iii) are satisfied regardless of the absolute value of $B$, since $\chi_\gamma^{\rm Che}$ is a fixed parameter of the emission geometry and $\mathcal{E}_c^{\rm Che} \propto B_*^{-1}$ grows at lower fields. Although the absolute departure of the refractive index $\Delta n$ is significantly small at lower fields ($B\sim10^{10}-10^{11}$~G), the mechanism remains self-consistent. The required threshold for Cherenkov emission, $\gamma_{\rm Che}\propto \Delta n^{-1/2}$, is compensated by the high potential drops achievable in short-period pulsars, which naturally provide the necessary Lorentz factors for these objects to lie above the QC death line. Indeed, for $B \sim 10^{11}$~G the death line requires $P \lesssim 1$~s, while for $B \sim 10^{10}$~G one needs $P \lesssim 10^{-1}$~s. In this regime, shorter rotation periods boost the maximum Lorentz factor accessible in the vacuum gap, $\gamma_{\rm max} =e\Delta\Phi_{\rm max}/m_e c^2\approx  10^6\,P^{-2} B_*$. The net effect is that the condition $\gamma_{\rm max} > \gamma_{\rm Che}$ can be satisfied above the QC death line even at low fields, confirming the formal self-consistency of the mechanism in this regime. The upper boundary of validity, $B\lesssim 10^{17}$~G, is set independently by the condition $\Delta n_{\parallel,\perp} \lesssim 1$, beyond which the weak-field expansion of the Euler-Heisenberg Lagrangian breaks down and the present derivation of the refractive index loses reliability.

Regarding the multi-wavelength implications, the QC mechanism reshapes the vacuum gap structure, leading to distinct observational predictions. First, the modified gap height interferes with the back-streaming particle flux, potentially maintaining polar cap temperatures detectable in the thermal X-ray band even for long-period pulsars near the traditional death line \cite{2025Natur.642..583W}. Second, high-energy secondary cascades are structurally altered: since primary QC photons are emitted strongly peaked at a characteristic quantum parameter $\chi_\gamma^{\rm Che} \approx 0.75$, the resulting spectrum lacks the broad power-law tail typical of curvature radiation. This implies a distinct, sharply peaked (nearly monoenergetic) signature in the hard X-ray to soft gamma-ray band prior to cascade reprocessing. These distinct gamma-ray signatures could be probed by next-generation observatories such as CTAO \cite{ctao}. Finally, a critical aspect of extending the radio population to ultra-long periods is detectability. Pulsars operating near the QC boundary exhibit inherently low spin-down luminosities. However, for a hypothetical pulsar just above the QC death line (e.g., $P = 10$~s, $\dot{P} = 10^{-14}$ s s$^{-1}$), the available spin-down energy is $L_{\rm sd} \sim 4 \times 10^{29}$~erg s$^{-1}$. Assuming a typical radio emission efficiency of $\eta_{\rm rad} \sim 10^{-4}$, the expected radio luminosity is $L_{\rm rad} \sim 4 \times 10^{25}$~erg s$^{-1}$, which translates to a flux density on the order of $\sim 10$--$100$~$\mu$Jy for a source at 1~kpc. While faint, this predicted flux provides a clear, falsifiable target that falls well within the sensitivity limits of next-generation radio facilities such as the Square Kilometre Array (SKA), which is projected to reach sub-$\mu$Jy sensitivities in deep integrations, and the next-generation Very Large Array (ngVLA).

\section{Conclusion}\label{sec5}

In summary, Cherenkov radiation from the quantum vacuum constitutes a viable pair-creation channel in NS polar caps, extending the paradigm of pulsar radio emission. Our modeling of the electrodynamic gap shows that Cherenkov photons are efficiently produced at low altitudes, with mean free paths that are significantly shorter than the gap height ($\ell_{\rm ph}\ll \ell_{\rm acc} \sim h_{\rm gap}$). This ensures rapid conversion into electron-positron pairs even in regimes where CR and/or RICS are suppressed. Consequently, our analysis reveals a Cherenkov-enabled death line that extends the viable parameter space for radio-emitting pulsars. This framework confirms that sustained radio emission remains physically possible for some pulsars within/below the CR and/or RICS death valleys. Although the nature of individual long-period transients such as GLEAM-X~J1627--5235 and GPM~J1839--10 remains under investigation, our work demonstrates that NSs can, in principle, sustain radio emission in this long period domain. This result, together with the possibility that some long-period transients could be magnetars that give rise to fast radio bursts (FRBs) \cite{2025ApJ...990L..49D}, suggests the interesting possibility of a unified picture connecting long-period transients, magnetars, and FRBs \cite{rea2026}. More broadly, our results confirm NSs as natural laboratories for strong-field QED, where vacuum polarization fundamentally reshapes magnetospheric physics. The predicted population of Cherenkov-powered pulsars represents a clear target for next-generation radio surveys, offering an opportunity to test non-linear QED in the astrophysical environment.

\begin{acknowledgments}
The authors are grateful to the anonymous reviewers for their constructive comments and valuable suggestions, which significantly improved the quality and clarity of this manuscript. We also thank the Extreme Light Infrastructure (ELI) ERIC and the ELI Beamlines Facility for the hospitality, infrastructure, and valuable discussions during the development of this work. This study was financed in part by the Coordena\c{c}\~ao de Aperfei\c{c}oamento de Pessoal de N\'ivel Superior - Brasil (CAPES) - Finance Code 001, through the PDSE program (Grant No. 06/2024). T.A.O.G. acknowledges financial support from FAPES (grants No. 039/2023 and 0022/2024). J.G.C. is grateful for the support of FAPES (1020/2022, 1081/2022, 976/2022, 332/2023), CNPq (306018/2025-0), and FAPESP (grant No. 2021/01089-1). This work was also supported by the Universidade Federal do Esp\'irito Santo (UFES).
\end{acknowledgments}

\section*{Declarations}
\begin{itemize}
    \item \textbf{Author contributions}\\
    T.A.O.G. formulated the original idea, derived the equations, performed the numerical analysis, generated the figures, and wrote the main manuscript. J.G.C. and J.A.R. supervised the astrophysical framing and the application of the theoretical model to pulsar magnetospheres. S.V.B. supervised the theoretical development regarding non-linear electrodynamics and strong-field QED. M.J. and A.J.M. contributed to the theoretical discussions and analytical formulations of QED in strong magnetic fields. All authors discussed the results, reviewed the manuscript, and agreed on its final version.
    
    \item \textbf{Funding}\\
    This work was supported by the Funda\c{c}\~ao de Amparo \`a Pesquisa e Inova\c{c}\~ao do Esp\'irito Santo (FAPES) and the Universidade Federal do Esp\'irito Santo (UFES). Specific grant numbers are detailed in the Acknowledgments.

    \item \textbf{Conflict of interest}\\
    The authors declare no competing interests.

    \item \textbf{Data availability}\\
    The observational datasets analyzed in this study are publicly available in the ATNF Pulsar Catalogue \cite{atnf} and the McGill Magnetar Catalog \cite{mcgill2014}. The source data for Figure~\ref{fig:dlnew} is provided as Supplementary Data. Figure~\ref{fig:figura} is a schematic diagram generated directly in LaTeX and contains no underlying numerical data. Any additional data are available upon request from the corresponding author, Tales A. O. Gomes (\href{mailto:tales.gomes@edu.ufes.br}{tales.gomes@edu.ufes.br}).

    \item \textbf{Code availability}\\
    The custom scripts used to calculate the strong-field QED parameters and generate the theoretical boundaries in Figure~\ref{fig:dlnew} are available as Supplementary Software. The base code for plotting the observational transient background in Figure~\ref{fig:dlnew} is open-source and available at \url{https://github.com/nhurleywalker/GPMTransient}. Any additional code details are available upon request from the corresponding author, Tales A. O. Gomes (\href{mailto:tales.gomes@edu.ufes.br}{tales.gomes@edu.ufes.br}).
\end{itemize}

\bibliographystyle{apsrev4-2}
\bibliography{biblio}

@article{Sauter1931,
  author  = {Sauter, Fritz},
  title   = {{{\"U}ber das Verhalten eines Elektrons im homogenen elektrischen Feld nach der relativistischen Theorie Diracs}},
  journal = {Z. Phys.},
  year    = {1931},
  volume  = {69},
  number  = {11-12},
  pages   = {742--764},
  doi     = {10.1007/BF01339461}
}

@article{EulerHeisenberg1936,
  author  = {Heisenberg, W. and Euler, H.},
  title   = {{Folgerungen aus der Diracschen Theorie des Positrons}},
  journal = {Z. Phys.},
  year    = {1936},
  volume  = {98},
  number  = {11-12},
  pages   = {714--732},
  doi     = {10.1007/BF01343663}
}

@article{Frank1937,
  author  = {Frank, I. M. and Tamm, I. E.},
  title   = {{Coherent visible radiation of fast electrons passing through matter}},
  journal = {C. R. Acad. Sci. URSS},
  volume  = {14},
  number  = {3},
  pages   = {109--114},
  year    = {1937}
}

@article{schwinger1949,
  title   = {On the classical radiation of accelerated electrons},
  author  = {Schwinger, Julian},
  journal = {Phys. Rev.},
  volume  = {75},
  number  = {12},
  pages   = {1912},
  year    = {1949}
}

@article{schwinger1951,
  title   = {On gauge invariance and vacuum polarization},
  author  = {Schwinger, Julian},
  journal = {Phys. Rev.},
  volume  = {82},
  number  = {5},
  pages   = {664--679},
  year    = {1951},
  doi     = {10.1103/PhysRev.82.664}
}

@phdthesis{toll1952,
  author = {Toll, John Sampson},
  title  = {{The Dispersion Relation for Light and its Application to Problems Involving Electron Pairs}},
  school = {Princeton University},
  year   = {1952}
}

@article{Erber1966,
  author  = {Erber, Thomas},
  title   = {High-energy electromagnetic conversion processes in intense magnetic fields},
  journal = {Rev. Mod. Phys.},
  year    = {1966},
  volume  = {38},
  number  = {4},
  pages   = {626--659},
  doi     = {10.1103/RevModPhys.38.626}
}

@article{GoldreichJulian1969,
  author  = {Goldreich, Peter and Julian, William H.},
  title   = {Pulsar electrodynamics},
  journal = {Astrophys. J.},
  year    = {1969},
  volume  = {157},
  pages   = {869},
  doi     = {10.1086/150119}
}

@article{ADLER1971,
  title   = {Photon splitting and photon dispersion in a strong magnetic field},
  author  = {Adler, Stephen L.},
  journal = {Ann. Phys.},
  volume  = {67},
  number  = {2},
  pages   = {599--647},
  year    = {1971},
  doi     = {10.1016/0003-4916(71)90154-0}
}

@article{Sturrock1971,
  author  = {Sturrock, P. A.},
  title   = {A model of pulsars},
  journal = {Astrophys. J.},
  year    = {1971},
  volume  = {164},
  pages   = {529},
  doi     = {10.1086/150865}
}

@article{RITUS1972,
  title   = {Radiative corrections in quantum electrodynamics with intense field and their analytical properties},
  author  = {Ritus, V. I.},
  journal = {Ann. Phys.},
  volume  = {69},
  number  = {2},
  pages   = {555--582},
  year    = {1972},
  doi     = {10.1016/0003-4916(72)90191-1}
}

@article{Ruderman1975,
  author  = {Ruderman, M. A. and Sutherland, P. G.},
  title   = {Theory of pulsars: polar gaps, sparks, and coherent microwave radiation},
  journal = {Astrophys. J.},
  year    = {1975},
  volume  = {196},
  pages   = {51--72},
  doi     = {10.1086/153393}
}

@article{Schwinger1976,
  author  = {Schwinger, Julian and Tsai, Wuyang and Erber, Thomas},
  title   = {Classical and quantum theory of synergic synchrotron-{C}herenkov radiation},
  journal = {Ann. Phys.},
  year    = {1976},
  volume  = {96},
  number  = {2},
  pages   = {303--332},
  doi     = {10.1016/0003-4916(76)90194-9}
}

@article{narozhny1979,
  title   = {Radiation corrections to quantum processes in an intense electromagnetic field},
  author  = {Narozhny, N. B.},
  journal = {Phys. Rev. D},
  volume  = {20},
  number  = {6},
  pages   = {1313--1320},
  year    = {1979},
  doi     = {10.1103/PhysRevD.20.1313}
}

@inbook{shapiro83,
  author    = {Shapiro, Stuart L. and Teukolsky, Saul A.},
  title     = {Black Holes, White Dwarfs, and Neutron Stars: The Physics of Compact Objects},
  publisher = {John Wiley \& Sons, Ltd},
  year      = {1983},
  chapter   = {10},
  pages     = {267--305},
  address   = {New York},
  doi       = {10.1002/9783527617661.ch10}
}

@article{ritus1985quantum,
  title   = {Quantum effects of the interaction of elementary particles with an intense electromagnetic field},
  author  = {Ritus, V. I.},
  journal = {J. Sov. Laser Res.},
  volume  = {6},
  number  = {5},
  year    = {1985}
}

@article{chen1993pulsar,
  title   = {Pulsar death lines and death valley},
  author  = {Chen, Kaiyou and Ruderman, Malvin},
  journal = {Astrophys. J.},
  volume  = {402},
  pages   = {264--270},
  year    = {1993}
}

@article{thompson1996,
  author  = {Thompson, Christopher and Duncan, Robert C.},
  title   = {{The soft gamma repeaters as very strongly magnetized neutron stars. II. {Q}uiescent neutrino, {X}-ray, and {A}lfven wave emission}},
  journal = {Astrophys. J.},
  year    = {1996},
  volume  = {473},
  pages   = {322},
  doi     = {10.1086/178147}
}

@article{Bamber1996,
  title   = {Studies of nonlinear {QED} in collisions of 46.6 {G}eV electrons with intense laser pulses},
  author  = {Bamber, C. and others},
  journal = {Phys. Rev. D},
  volume  = {60},
  number  = {9},
  pages   = {092004},
  year    = {1999},
  doi     = {10.1103/PhysRevD.60.092004}
}

@article{bula1996,
  author  = {Bula, C. and others},
  title   = {Observation of nonlinear effects in {C}ompton scattering},
  journal = {Phys. Rev. Lett.},
  volume  = {76},
  pages   = {3116--3119},
  year    = {1996},
  doi     = {10.1103/PhysRevLett.76.3116}
}

@inproceedings{Burke1998,
  author    = {Burke, D. L. and others},
  title     = {Quantum aspects, experimental results, and beam physics implications of {E}-144},
  booktitle = {15th Advanced ICFA Beam Dynamics Workshop on Quantum Aspects of Beam Physics},
  pages     = {357--368},
  year      = {1998}
}

@article{Hibschman_2001,
doi = {10.1086/321378},
url = {https://doi.org/10.1086/321378},
year = {2001},
month = {jun},
publisher = {},
volume = {554},
number = {1},
pages = {624},
author = {Hibschman, Johann A. and Arons, Jonathan},
title = {Pair Multiplicities and Pulsar Death},
journal = {The Astrophysical Journal},
}

@article{BET2003,
  title   = {Light intensification towards the {S}chwinger limit},
  author  = {Bulanov, Sergei V. and Esirkepov, Timur and Tajima, Toshiki},
  journal = {Phys. Rev. Lett.},
  volume  = {91},
  number  = {8},
  pages   = {085001},
  year    = {2003}
}

@article{atnf,
  author  = {Manchester, R. N. and Hobbs, G. B. and Teoh, A. and Hobbs, M.},
  title   = {{The {A}ustralia {T}elescope {N}ational {F}acility {P}ulsar {C}atalogue}},
  journal = {Astron. J.},
  year    = {2005},
  volume  = {129},
  number  = {4},
  pages   = {1993--2006},
  doi     = {10.1086/428488}
}

@article{MTB2006,
  title   = {Optics in the relativistic regime},
  author  = {Mourou, Gerard A. and Tajima, Toshiki and Bulanov, Sergei V.},
  journal = {Rev. Mod. Phys.},
  volume  = {78},
  number  = {2},
  pages   = {309--371},
  year    = {2006}
}

@article{deluca2006,
  author  = {De Luca, A. and Caraveo, P. A. and Mereghetti, S. and Tiengo, A. and Bignami, G. F.},
  title   = {A long-period, violently variable {X}-ray source in a young supernova remnant},
  journal = {Science},
  year    = {2006},
  volume  = {313},
  number  = {5788},
  pages   = {814--817},
  doi     = {10.1126/science.1129185}
}

@article{timokhin2010,
  author  = {Timokhin, A. N.},
  title   = {Time-dependent pair cascades in magnetospheres of neutron stars – {I}. {D}ynamics of the polar cap cascade with no particle supply from the neutron star surface},
  journal = {Mon. Not. R. Astron. Soc.},
  volume  = {408},
  number  = {4},
  pages   = {2092--2114},
  year    = {2010},
  doi     = {10.1111/j.1365-2966.2010.17286.x}
}

@article{baring2011,
doi = {10.1088/0004-637X/733/1/61},
url = {https://doi.org/10.1088/0004-637X/733/1/61},
year = {2011},
month = {may},
publisher = {The American Astronomical Society},
volume = {733},
number = {1},
pages = {61},
author = {Baring, Matthew G. and Wadiasingh, Zorawar and Gonthier, Peter L.},
title = {COOLING RATES FOR RELATIVISTIC ELECTRONS UNDERGOING COMPTON SCATTERING IN STRONG MAGNETIC FIELDS},
journal = {The Astrophysical Journal},
}

@article{timokhin2013,
  author  = {Timokhin, A. N. and Arons, J.},
  title   = {Current flow and pair creation at low altitude in rotation-powered pulsars' force-free magnetospheres: space charge limited flow},
  journal = {Mon. Not. R. Astron. Soc.},
  volume  = {429},
  number  = {1},
  pages   = {20--54},
  year    = {2013},
  doi     = {10.1093/mnras/sts298}
}

@inproceedings{ELIBeamlinesFacility,
  title     = {{ELI} {E}xtreme {L}ight {I}nfrastructure science and technology with ultra-intense lasers},
  author    = {Korn, Georg and LeGarrec, Bruno and Rus, Bedrich},
  booktitle = {CLEO: Science and Innovations},
  pages     = {CTu2D--7},
  year      = {2013}
}

@article{mcgill2014,
  author  = {Olausen, S. A. and Kaspi, V. M.},
  title   = {{The {McGill} {M}agnetar {C}atalog}},
  journal = {Astrophys. J. Suppl. Ser.},
  year    = {2014},
  volume  = {212},
  number  = {1},
  pages   = {6},
  doi     = {10.1088/0067-0049/212/1/6}
}

@article{rea2016,
  author  = {Rea, N. and others},
  title   = {Magnetar-like activity from the central compact object in the {SNR} {RCW103}},
  journal = {Astrophys. J. Lett.},
  year    = {2016},
  volume  = {828},
  number  = {1},
  pages   = {L13},
  doi     = {10.3847/2041-8205/828/1/L13}
}

@article{Aaboud2017,
  author  = {{ATLAS Collaboration}},
  title   = {Evidence for light-by-light scattering in heavy-ion collisions with the {ATLAS} detector at the {LHC}},
  journal = {Nat. Phys.},
  volume  = {13},
  number  = {9},
  pages   = {852--858},
  year    = {2017},
  doi     = {10.1038/nphys4208}
}

@article{Tan_2018,
  doi          = {10.3847/1538-4357/aade88},
  year         = {2018},
  month        = oct,
  volume       = {866},
  number       = {1},
  pages        = {54},
  author       = {Tan, C. M. and Bassa, C. G. and Cooper, S. and Dijkema, T. J. and Esposito, P. and Hessels, J. W. T. and Kondratiev, V. I. and Kramer, M. and Michilli, D. and Sanidas, S. and Shimwell, T. W. and Stappers, B. W. and van Leeuwen, J. and Cognard, I. and Grießmeier, J.-M. and Karastergiou, A. and Keane, E. F. and Sobey, C. and Weltevrede, P.},
  title        = {{LOFAR} Discovery of a 23.5 s Radio Pulsar},
  journal      = {Astrophys. J.}
}

@article{Macleod2019,
  author  = {Macleod, Alexander J. and Noble, Adam and Jaroszynski, Dino A.},
  title   = {{Cherenkov radiation from the quantum vacuum}},
  journal = {Phys. Rev. Lett.},
  year    = {2019},
  volume  = {122},
  number  = {16},
  pages   = {161601},
  doi     = {10.1103/PhysRevLett.122.161601}
}

@article{Bulanov2019,
  author  = {Bulanov, S. V. and Sasorov, P. and Bulanov, S. S. and Korn, G.},
  title   = {Synergic {C}herenkov-{C}ompton radiation},
  journal = {Phys. Rev. D},
  year    = {2019},
  volume  = {100},
  number  = {1},
  pages   = {016012},
  doi     = {10.1103/PhysRevD.100.016012}
}

@book{ctao,
    author = {{The CTA Consortium}},
    title = {Science with the Cherenkov Telescope Array},
    publisher = {World Scientific},
    address = {Singapore},
    year = {2019},
    doi = {10.1142/10986},
    URL = {https://www.worldscientific.com/doi/abs/10.1142/10986}
}

@inproceedings{APOLLON,
  title        = {{APOLLON}: a multi-{PW} laser facility},
  author       = {Jacquemot, Sylvie and Zeitoun, Philippe},
  booktitle    = {X-Ray Lasers and Coherent X-Ray Sources: Development and Applications XIII},
  volume       = {11111},
  pages        = {18},
  year         = {2019},
  organization = {SPIE}
}

@article{atlas2019,
  title   = {Observation of light-by-light scattering in ultraperipheral {Pb}+{Pb} collisions with the {ATLAS} detector},
  author  = {{ATLAS Collaboration}},
  journal = {Phys. Rev. Lett.},
  volume  = {123},
  number  = {5},
  pages   = {052001},
  year    = {2019},
  doi     = {10.1103/PhysRevLett.123.052001}
}

@article{philipov2020,
  title   = {Origin of pulsar radio emission},
  author  = {Philippov, Alexander and Timokhin, Andrey and Spitkovsky, Anatoly},
  journal = {Phys. Rev. Lett.},
  volume  = {124},
  number  = {24},
  pages   = {245101},
  year    = {2020},
  doi     = {10.1103/PhysRevLett.124.245101}
}

@article{EJLLI20201,
  title   = {The {PVLAS} experiment: A 25 year effort to measure vacuum magnetic birefringence},
  journal = {Phys. Rep.},
  volume  = {871},
  pages   = {1--74},
  year    = {2020},
  doi     = {10.1016/j.physrep.2020.06.001},
  author  = {Ejlli, A. and {Della Valle}, F. and Gastaldi, U. and Messineo, G. and Pengo, R. and Ruoso, G. and Zavattini, G.}
}

@article{SEL,
  title   = {Overview and status of station of extreme light toward 100 {PW}},
  author  = {Peng, Yujie and others},
  journal = {Rev. Laser Eng.},
  volume  = {49},
  number  = {2},
  pages   = {93},
  year    = {2021}
}

@article{caleb2022,
  author       = {Caleb, Manisha and Heywood, Ian and Rajwade, Kaustubh and Malenta, Mateusz and Stappers, Benjamin Willem and Barr, Ewan and Chen, Weiwei and Morello, Vincent and Sanidas, Sotiris and van den Eijnden, Jakob and Kramer, Michael and Buckley, David and Brink, Jaco and Motta, Sara Elisa and Woudt, Patrick and Weltevrede, Patrick and Jankowski, Fabian and Surnis, Mayuresh and Buchner, Sarah and Bezuidenhout, Mechiel Christiaan and Driessen, Laura Nicole and Fender, Rob},
  title        = {{Discovery of a radio-emitting neutron star with an ultra-long spin period of 76 s}},
  journal      = {Nat. Astron.},
  year         = {2022},
  month        = may,
  volume       = {6},
  pages        = {828--836},
  doi          = {10.1038/s41550-022-01688-x},
  archivePrefix= {arXiv},
  eprint       = {2206.01346},
  primaryClass = {astro-ph.HE}
}

@ARTICLE{loeb2022,
       author = {{Loeb}, Abraham and {Maoz}, Dan},
        title = "{A Hot Subdwarf Model for the 18.18 minutes Pulsar GLEAM-X}",
      journal = {Research Notes of the American Astronomical Society},
         year = 2022,
        month = feb,
       volume = {6},
       number = {2},
          eid = {27},
        pages = {27},
          doi = {10.3847/2515-5172/ac52f1},
archivePrefix = {arXiv},
       eprint = {2202.04949},
 primaryClass = {astro-ph.SR},
       adsurl = {https://ui.adsabs.harvard.edu/abs/2022RNAAS...6...27L}
}

@ARTICLE{katz2022,
       author = {{Katz}, J.~I.},
        title = "{GLEAM-X J162759.5{\ensuremath{-}}523504.3 as a white dwarf pulsar}",
      journal = {Astrophys. Space Sci},
         year = 2022,
        month = nov,
       volume = {367},
       number = {11},
          eid = {108},
        pages = {108},
          doi = {10.1007/s10509-022-04146-2},
archivePrefix = {arXiv},
       eprint = {2203.08112},
 primaryClass = {astro-ph.SR},
       adsurl = {https://ui.adsabs.harvard.edu/abs/2022Ap&SS.367..108K}
}

@article{2023Natur.619..487H,
  author       = {Hurley-Walker, N. and Rea, N. and McSweeney, S. J. and Meyers, B. W. and Lenc, E. and Heywood, I. and Hyman, S. D. and Men, Y. P. and Clarke, T. E. and {Coti Zelati}, F. and Price, D. C. and Horv{\'a}th, C. and Galvin, T. J. and Anderson, G. E. and Bahramian, A. and Barr, E. D. and Bhat, N. D. R. and Caleb, M. and Dall'Ora, M. and de Martino, D. and Giacintucci, S. and Morgan, J. S. and Rajwade, K. M. and Stappers, B. and Williams, A.},
  title        = {{A long-period radio transient active for three decades}},
  journal      = {Nature},
  year         = {2023},
  month        = jul,
  volume       = {619},
  number       = {7970},
  pages        = {487--490},
  doi          = {10.1038/s41586-023-06202-5},
  archivePrefix= {arXiv},
  eprint       = {2503.08036},
  primaryClass = {astro-ph.HE}
}

@article{2024NatAs...8.1159C,
  author       = {Caleb, M. and Lenc, E. and Kaplan, D. L. and Murphy, T. and Men, Y. P. and Shannon, R. M. and Ferrario, L. and Rajwade, K. M. and Clarke, T. E. and Giacintucci, S. and Hurley-Walker, N. and Hyman, S. D. and Lower, M. E. and McSweeney, Sam and Ravi, V. and Barr, E. D. and Buchner, S. and Flynn, C. M. L. and Hessels, J. W. T. and Kramer, M. and Pritchard, J. and Stappers, B. W.},
  title        = {{An emission-state-switching radio transient with a 54-minute period}},
  journal      = {Nat. Astron.},
  year         = {2024},
  month        = sep,
  volume       = {8},
  pages        = {1159--1168},
  doi          = {10.1038/s41550-024-02277-w},
  archivePrefix= {arXiv},
  eprint       = {2407.12266},
  primaryClass = {astro-ph.HE}
}

@article{NSFOPAL,
  title     = {Ultrafast laboratory astrophysics and planetary physics with {NSF}-{OPAL}},
  author    = {Polsin, Danae and others},
  journal   = {Bull. Am. Phys. Soc.},
  year      = {2024},
  publisher = {APS}
}

@misc{cooper2024,
      title={Beyond the Rotational Deathline: Radio Emission from Ultra-long Period Magnetars}, 
      author={A. J. Cooper and Z. Wadiasingh},
      year={2024},
      eprint={2406.04135},
      archivePrefix={arXiv},
      primaryClass={astro-ph.HE},
      url={https://arxiv.org/abs/2406.04135}, 
}

@article{2025ApJ...990L..49D,
  author       = {Dong, Fengqiu Adam and Clarke, Tracy E. and Curtin, Alice and Kumar, Ajay and Mckinven, Ryan and Shin, Kaitlyn and Stairs, Ingrid and Brar, Charanjot and Burdge, Kevin and Chatterjee, Shami and Cook, Amanda M. and Fonseca, Emmanuel and Gaensler, B. M. and Hessels, Jason W. and Kaspi, Victoria M. and Lazda, Mattias and Main, Robert and Masui, Kiyoshi W. and McKee, James W. and Meyers, Bradley W. and Pearlman, Aaron B. and Ransom, Scott M. and Scholz, Paul and Smith, Kendrick M. and Tan, Chia Min},
  title        = {{CHIME/Fast Radio Burst/Pulsar Discovery of a Nearby Long-period Radio Transient with a Timing Glitch}},
  journal      = {Astrophys. J.},
  year         = {2025},
  month        = sep,
  volume       = {990},
  number       = {2},
  eid          = {L49},
  pages        = {L49},
  doi          = {10.3847/2041-8213/adfa8e},
  archivePrefix= {arXiv},
  eprint       = {2407.07480},
  primaryClass = {astro-ph.HE}
}

@article{wang2025,
  author       = {Wang, Yuanming and Uttarkar, Pavan A. and Shannon, Ryan M. and Lee, Yu Wing Joshua and Dobie, Dougal and Wang, Ziteng and Bannister, Keith W. and Caleb, Manisha and Deller, Adam T. and Glowacki, Marcin and Jahns-Schindler, Joscha N. and Murphy, Tara and Anna-Thomas, Reshma and Bhat, N. D. R. and Deng, Xinping and Gupta, Vivek and Jaini, Akhil and James, Clancy W. and Tuthill, John},
  title        = {{The Discovery of a 41 s Radio Pulsar PSR J0311+1402 with ASKAP}},
  journal      = {Astrophys. J. Lett.},
  year         = {2025},
  month        = apr,
  volume       = {982},
  number       = {2},
  eid          = {L53},
  pages        = {L53},
  doi          = {10.3847/2041-8213/adbe61},
  archivePrefix= {arXiv},
  eprint       = {2503.07936},
  primaryClass = {astro-ph.HE}
}

@ARTICLE{2025Natur.642..583W,
       author = {{Wang}, Ziteng and {Rea}, Nanda and {Bao}, Tong and {Kaplan}, David L. and {Lenc}, Emil and {Wadiasingh}, Zorawar and {Hare}, Jeremy and {Zic}, Andrew and {Anumarlapudi}, Akash and {Bera}, Apurba and {Beniamini}, Paz and {Cooper}, A.~J. and {Clarke}, Tracy E. and {Deller}, Adam T. and {Dawson}, J.~R. and {Glowacki}, Marcin and {Hurley-Walker}, Natasha and {McSweeney}, S.~J. and {Polisensky}, Emil J. and {Peters}, Wendy M. and {Younes}, George and {Bannister}, Keith W. and {Caleb}, Manisha and {Dage}, Kristen C. and {James}, Clancy W. and {Kasliwal}, Mansi M. and {Karambelkar}, Viraj and {Lower}, Marcus E. and {Mori}, Kaya and {Ocker}, Stella Koch and {P{\'e}rez-Torres}, Miguel and {Qiu}, Hao and {Rose}, Kovi and {Shannon}, Ryan M. and {Taub}, Rhianna and {Wang}, Fayin and {Wang}, Yuanming and {Zhao}, Zhenyin and {Bhat}, N.~D. Ramesh and {Dobie}, Dougal and {Driessen}, Laura N. and {Murphy}, Tara and {Jaini}, Akhil and {Deng}, Xinping and {Jahns-Schindler}, Joscha N. and {Lee}, Y.~W. Joshua and {Pritchard}, Joshua and {Tuthill}, John and {Thyagarajan}, Nithyanandan},
        title = "{Detection of X-ray emission from a bright long-period radio transient}",
      journal = {Natur.},
         year = 2025,
        month = jun,
       volume = {642},
       number = {8068},
        pages = {583-586},
          doi = {10.1038/s41586-025-09077-w},
}

@misc{rea2025magnetars,
  title        = {Magnetars},
  author       = {Rea, Nanda and {De Grandis}, Davide},
  year         = {2025},
  eprint       = {2503.04442},
  archivePrefix= {arXiv},
  primaryClass = {astro-ph.HE},
  url          = {https://arxiv.org/abs/2503.04442}
}

@misc{rea2026,
  title        = {Long Period Transients ({LPTs}): a comprehensive review},
  author       = {Rea, Nanda and Hurley-Walker, Natasha and Caleb, Manisha},
  year         = {2026},
  eprint       = {2601.10393},
  archivePrefix= {arXiv},
  primaryClass = {astro-ph.HE},
  url          = {https://arxiv.org/abs/2601.10393}
}

\end{document}